\documentclass[twocolumn]{autart}    

\usepackage{graphicx}      
\usepackage[square,sort,comma,numbers]{natbib}

\usepackage{amsmath,amsfonts,amssymb}
\usepackage{tikz}
\usepackage{subfigure}
\usepackage{paralist}
\usepackage{pgfplots}
\usetikzlibrary{arrows,shapes,trees}
\newtheorem{Theorem}{Theorem} 
\newtheorem{Corollary}{Corollary} 
\newtheorem{Lemma}{Lemma} 
\newtheorem{Remark}{Remark} 

\newcommand{\mc}[1]{\mathcal{#1}}

\newcommand{\ik}{^{(k)}}

\usepackage{tikz}
\usepackage{pgfplots}
\usepackage{arydshln}
\usetikzlibrary{arrows,shapes,trees}

\usepackage{nicefrac} 

\newcommand{\tild}[1]{\widetilde{#1}}

\begin{document}

\begin{frontmatter}

\title{Distributed Nonlinear Observer with Robust Performance \\ - A Circle Criterion Approach} 

\thanks[footnoteinfo]{This work was supported
by the German
Research Foundation (DFG) through
the Cluster of Excellence in Simulation Technology (EXC 310/1) at the
University of Stuttgart.}

\author{Jingbo Wu},
\author{Frank Allg\"ower}

\address{Institute for Systems Theory and Automatic Control, University of Stuttgart,
   70550 Stuttgart, Germany (e-mail: \{jingbo.wu, allgower\}@ ist.uni-stuttgart.de)}

\begin{abstract}                
In this paper, we present a distributed version of the KYP-Lemma with the goal to express the strictly positive real-property for a class of physically interconnected systems by a set of local LMI-conditions. The resulting conditions are subsequently used to constructively design distributed circle criterion estimators, which are able to collectively estimate an underlying linear system with a sector bounded nonlinearity.
\end{abstract}


\end{frontmatter}




\section{Introduction}

Estimator design has been an essential part of controller design ever since
the development of state-space based controllers. Milestones were laid by
the Luenberger Observer \cite{Luenberger1966}, the Kalman Filter \cite{Kalman1960}, and the $\mathcal{H}_\infty$-Filter \cite{Shaked1990}. 

While in the classical estimator design one estimator is used for one
system, designing distributed estimators have gained attention since a
distributed Kalman Filter was presented in \cite{OlfatiSaber:2005vm},
\cite{OlfatiSaber2007}, \cite{Carli2008}. In a distributed estimator setup, multiple
estimators create an estimate of the system's state, while cooperating
with each other. In this setup, even when every single estimator may be
able to obtain an estimate of the state on its own, cooperation reduces the
effects of model and measurement disturbances
\cite{SS-2009}.
Also, the situations are not uncommon where every single estimator is unable
to obtain an estimate of the state on its own and cooperation becomes an
essential prerequisite \cite{Ugrinovskii2011},
\cite{Ugrinovskii2011a}. 

Where the literature review above shows that there is a considerable number of results to address distributed estimation for linear systems, nonlinear systems have barely been considered. When looking at existing nonlinear estimation algorithms in literature, one can notice that many of them require some kind of transformation upfront. For instance, the Extended Luenberger observer \cite{Zeitz1987} and the High-gain Observer \cite{Khalil2014} require a transformation to observability normal form. However, in the case when multiple sensing units cooperate in a distributed setup, a transformation of coordinate hinders the efficient exchange of information, unless the transformed coordinates are the same. Restricting the state transformation to be the same for all sensing units however requires the measured information to be essentially the same, which is a trivial case. On the other hand, without coordinate transformation, there are observer design methods in literature that deal with systems described by a linear state space model with an additive, sector-bounded nonlinearity \citep{Arcak2001},\cite{Acikmese2011}. 

In this paper, we aim at extending LMI-based methods for distributed estimation such as \cite{Ugrinovskii2011} in order to deal with linear systems with an additive nonlinearity. Besides for globally Lipschitz nonlinearities, we will mainly present a design approach for a distributed circle criterion observer. This requires a distributed formulation of the KYP-Lemma, which shows us that the regular approach of taking the sum-of-squares Lyapunov function $V=\sum_{k=1}^N x_k P_x x_k$ as in \cite{Ugrinovskii2011} is not appropriate for this case. 

The rest of the paper is organized as follows: In Section II, we introduce the notation, some preliminaries on graph theory, and the respective system class. Then, in Section III we show an intuitive approach and a motivating example, where the sum-of-squares Lyapunov function $V=\sum_{k=1}^N x_k P_x x_k$ fails. This effect is subsequently discussed by the development of a distributed version of the KYP-Lemma in Section IV. Section V then deals with a generalized LMI-based construction method which overcomes the drawback of the intuitive approach. A simulation example is shown in Section VI.

\section{Preliminaries}

Throughout the paper the following notation is used: 
Let $A$ be a square matrix. If $A$ is positive definite, it is denoted
$A > 0$, and we write $A < 0$, if $A$ is negative definite. The norm of a matrix $\| A \|$ is defined as any induced matrix norm.


\subsection{Communication graphs}

In this section we summarize some notation from the graph theory.
We use undirected, unweighted graphs $\mc{G} = (\mc{V} , \mc{E})$ to describe the communication topology between the individual agents. 
$\mc{V} = \{v_1,...,v_N\}$ is the set of vertices, where $v_k \in \mc{V}$ represents the $k$-th agent. 
$\mc{E} \subseteq \mc{V} \times \mc{V}$ are the sets of edges, which model the information flow, i.e. the $k$-th agent can communicate with agent $j$ if and only if $(v_j,v_k) \in \mc{E}$. 
Since the graph is undirected, $(v_j,v_k) \in \mc{E}$ implies that $(v_k,v_j) \in \mc{E}$.
The set of vertices that agent $k$ receives information from is called the neighbourhood of agent $k$, which is denoted by $\mathcal{N}_k=\{j: (v_j,v_k) \in \mc{E}\}$. The degree $p_k$ of a vertex $k$ is defined as the number of vertices in $\mathcal{N}_k$. Assuming the graph as undirected is restrictive in general, however, we will later show that it is a sensible assumption for the problem of constructing the distributed circle criterion estimator.

\subsection{System model}
We consider the $n$-dimensional system
\vspace{-0.2cm}
\begin{equation}\label{sys:nonlinear}
\begin{aligned}
\dot{x} &= Ax + B_\phi \phi (H x) + B_\theta \theta ( \tild H x) + g(u) + B_w w  \\
y  &= Cx
\end{aligned}
\end{equation}
\vspace{-0.7cm}

where $x \in \mathbb{R}^{n}$ is the state variable, $u \in \mathbb{R}^{m}$ is the control input, $y \in \mathbb{R}^{q}$ is the output vector, and
$w(t) \in \mathbb{R}^l$ is an exogenous disturbance in the $\mathcal{L}_2$-space.
$\phi(\cdot)$ is a known $r$-dimensional nonlinearity satisfying \vspace{-0.2cm}
\begin{equation}\label{eq:monoton_nonlinearity}
\phi = \begin{bmatrix}
\phi_1(\sum_{j=1}^n H_{1j} x_j ) \\ \vdots \\ \phi_r (\sum_{j=1}^n H_{rj} x_j )
\end{bmatrix},
\end{equation} 
\vspace{-0.7cm} 

where every $\phi_i(\cdot)$ is a scalar nondecreasing function and
$\theta(\cdot)$ is a known $\tild r$-dimensional nonlinearity satisfying the incremental quadratic constraint
\begin{equation}\label{eq:qc-nonlinearity}
(a-b)^\top (a-b) \geq \tau^2  (\theta(a) - \theta(b))^\top (\theta(a) - \theta(b)),
\end{equation}
for any $a,b \in \mathbb{R}^{\tild r}$.
In fact, in many practical applications, the state $x(t)$ will be restricted to a bounded set $\mathcal{X}$. In this case, it suffices for \eqref{eq:monoton_nonlinearity}, \eqref{eq:qc-nonlinearity} to hold on this bounded set $\mathcal{X}$. 

System \eqref{sys:nonlinear} allows for both Lipschitz-nonlinearities and monotonous non-Lipschitz-nonlinearities, which together cover a large range of possible nonlinearities, similar to the incremental quadratic constraint \cite{Acikmese2011}
\vspace{-0.2cm}
\begin{equation*}
\begin{bmatrix}
a-b \\
\theta(a)-\theta(b)
\end{bmatrix}^\top
M
\begin{bmatrix}
a-b \\
\theta(a)-\theta(b)
\end{bmatrix}.
\end{equation*}
\vspace{-0.5cm}

\subsection{Problem statement}

The problem considered in this paper is to design $N$ local estimators for \eqref{sys:nonlinear}, where every local estimator $i=1,...,N$ relies only on the local measurement $y_k$ and communication with the neighboring estimators. In the following, we will denote 
\vspace{-0.2cm}
\begin{equation*}
y= \begin{bmatrix}
y_1 \\ y_2 \\ \vdots \\ y_N
\end{bmatrix} = \begin{bmatrix}
C_1 x \\ C_2 x \\ \vdots \\ C_N x 
\end{bmatrix}.
\end{equation*}
\vspace{-0.7cm}

The vector of local estimates will be denoted 
$\hat{x}_k \in \mathbb{R}^{n}$, and the local estimation error vector is defined as $e_k= x -\hat{x}_k$. The aggregated vector for all local estimation error vectors is denoted $e^\top=[e_1^\top, ..., e_N^\top]^\top$. Since the separation principle does not hold in general for nonlinear systems, we need to make a technical assumption on the closed loop system in order to avoid finite escape time \cite{Arcak2001}:


\noindent\textbf{Assumption 1}: Given initial conditions $x(0)$ and a control input $g(u)$, if $e(t) \in \mathcal{L}^e_\infty$, then $x(t) \in \mathcal{L}^e_\infty$.


Now, the distributed estimation problem can be expressed as following:

\textbf{Problem 1 (Distributed estimation):} Design a group of $N$ estimators with respective estimation $\hat{x}_k(t)$, $k=1, \ldots,N$, such that the following two properties are satisfied simultaneously:
 
\begin{compactenum}[(i)]
\item In the absence of disturbances (i.e., when
  $w=0$), the estimation errors decay so that $e_k \to 0$
  exponentially for all $k=1,...,N$. 

\item The estimators provide guaranteed
  $\mathcal{H}_{\infty}$ performance in the sense that 
\begin{equation}\label{eq:hinf-performance}
\begin{aligned}
 \sum_{k=1}^N \int_0^\infty e_k^\top W_k e_k  dt
\leq N \gamma^2 \| w \|^2_{\mathcal{L}_2} + I_0.
\end{aligned}
\end{equation}
$W_k$ is a positive semi-definite weighting matrix and $I_0$ is the cost due to the estimators' uncertainty about the initial conditions of the system.
\end{compactenum}


In particular, the estimators shall form a distributed setup in the way that the dynamics of each estimation $\hat{x}_k$ only depends on the local measurement $y_k$ and communication with the neighboring estimators $j$.

%
\section{An intuitive approach}
An intuitive approach to solve Problem 1 is by adapting the method introduced for estimation of linear systems in \cite{Ugrinovskii2011a},\cite{Wu2014}: There, for every estimator $k$, a respective LMI condition is derived, which allows for distributed calculation of the required filter gains \cite{Wu2015b}.

These LMI-conditions can be extended with respect to \eqref{eq:qc-nonlinearity} by adding the SPR-condition as done in \cite{Arcak2001}. The design conditions, which result from this intuitive approach are shown in the following.

\subsection{Design conditions}
We define the matrices
\begin{align*}
Q_k =& P_k A + A^\top P_k - G_k C_k - (G _k C_k )^\top - p_k F_k  - p_k F_k^\top \\ &  + \alpha P_k + p_k \pi_k  P_k,
\end{align*}
where $P_k \in \mathbb{R}^{\sigma_k \times \sigma_k}$ is a symmetric,
positive definite matrix. $\pi_k$ and $\alpha$ are positive 
constants which will later play the role of design parameters. 

Let the estimator dynamics be proposed as
\begin{equation}\label{sys:intuitive_estimator}
\begin{aligned}
\dot{\hat{x}}_k =& A \hat{x}_k + B_\phi \phi (H \hat{x}_k + \tild L_k(y_k - C_k \hat{x}_k) ) + B_\theta \theta(\tild H \hat x_k) \\ 
& +g(u) + L_k(y_k- C_k \hat{x}_k) + K_{k} \sum_{j \in \mathcal{N}_k} (\hat{x}_j- \hat{x}_k).
\end{aligned}
\end{equation}
Then, we have following design conditions.


\begin{Theorem}
Let a collection of matrices
$F_k$, $G_k$ and $P_k$, $k=1,\ldots,N$, be a solution of the LMIs 
\begin{equation}\label{LMI:intuitive}
\begin{aligned}
\!\!\left[\begin{array}{ccc;{2pt/2pt}ccc}
Q_k + \overline W_k & P_k B_\theta & P_k B_w  & F_k & \hdots & F_k   \\
(P_k B_\theta)^\top & -\tau^2 I & 0  & 0 & 0 & 0\\ 
(P_k B_w)^\top & 0 & -\gamma^2 I  & 0 & 0 & 0 \\
\hdashline[2pt/2pt] 
F_k ^\top  & 0 & 0  & - \pi_{j_1} \! P_{j_1} & 0 & 0\\
\vdots & 0 & 0  & 0 & \ddots & 0 \\
F_k^\top & 0 & 0  & 0 & 0 & -\pi_{j_{p_k}}  P_{j_{p_k}} 
\end{array}\right]
<  0
\end{aligned}
\end{equation}
and the equality constraints
\begin{equation}
- P_k B_\phi = H - \widetilde{L}_k C_k
\end{equation}
for all $k=1,...,N$, where $\overline{W}_k= W_k+ \tild H^\top \tild H$, then Problem 1 admits a solution of the form \eqref{sys:intuitive_estimator}, where
\begin{equation}\label{eq:intuitive_filtergains}
\begin{aligned}
L_k &= P_k^{-1} G_k \\
K_k &= P_k^{-1} F_k.
\end{aligned}
\end{equation}
\end{Theorem}
\begin{Remark}
In \eqref{LMI:intuitive}, the indexes $j_1,...,j_{p_k}$ enumerate the neighbors of estimator $k$. Strictly speaking, $j_1\ik,...,j_{p_k}\ik$ is required as notation, but in this paper, we drop the superscript $(k)$ to keep the notation simple.
\end{Remark}

The proof is omitted here because a more general version will be introduced and thoroughly proven later.
This approach works in some cases of Problem 1, but however has significant limitations due to conservativeness of the approach. One example, where is subsequently fails, is given in the following.

\subsection{A motivating example}

Consider the six-dimensional oscillator
\begin{equation} \label{sys:example_6dim_oszi}
\begin{aligned}
\dot{x}=&\begin{bmatrix}
0 & 1& 0& 1& 0& 1 \\
-1 &0& 1& 0& 1& 0 \\
0 &-1& 0& 1& 0& 1 \\
-1 &0& -1& 0& 1& 0 \\
0 &-1& 0& -1& 0& 1 \\
-1 &0& -1& 0& -1& 0
\end{bmatrix} x  \\
&+  B_\phi \phi(Hx) + B_w w +g(u)
\end{aligned}
\end{equation}
with the monotonously increasing nonlinearity $\phi(\cdot)$, where
$B_\phi=\begin{bmatrix}
1 & 0 & 0 & -1 & 0 & 0
\end{bmatrix}^\top$, 
$H=\begin{bmatrix}
1 & 1 & 1 & 1 & 1 & 1
\end{bmatrix}$, and $B_w=\begin{bmatrix}
1 & 1 & 1 & 1 & 1 & 1
\end{bmatrix}^\top$.
The individual measurements are
\begin{equation}\label{sys:example_6dim_oszi_output}
\begin{aligned}
y_1=x_2 \!-\! x_1,
&&
y_2=x_3 \!-\!  x_2, 
&&
y_3=x_4 \!-\!  x_3,\\
y_4=
x_5 \!-\! x_4,
&&
y_5=
x_6 \!-\! x_5,
&&
y_6=
x_1 \!-\! x_6,
\end{aligned}
\end{equation}
where $x=[x_1,...,x_6]^\top$, and let the estimator be connected by a ring-type communication topology $\mathcal{E}=\{(v_k, v_{k+1}),(v_{k+1}, v_k) | k=1,...,5\} \cup \{(v_6, v_1), (v_1, v_6)\}$.

Applying the solution method \eqref{sys:intuitive_estimator}-\eqref{eq:intuitive_filtergains} by using a numerical solver like \texttt{YALMIP/SEDUMI} will immediately yield infeasibility. In remainder of the paper, we will present the reason for this and introduce a more general approach in order to overcome this problem.

\section{A distributed KYP-Lemma}

In the single-system case, where we have the LTI system
\begin{equation}\label{sys:LTI}
\begin{aligned}
\dot x &= \tild Ax + Bu \\
y &=Ex,
\end{aligned}
\end{equation}
the closed loop of $\eqref{sys:LTI}$ with the nonlinear feedback $u=-\psi(y) \in \mathbb{R}^p$ satisfying $y^\top \psi(y) \geq 0$, is globally asymptotically stable, if the $p \times p$ transfer function matrix $G(s)=E(sI-\tild A)^{-1}B$ is strictly positive real (SPR). Further, from the Kalman-Yakubovich-Popov (KYP), we have following necessary and sufficient conditions.

\begin{Lemma}[Lemma 6.3 in \cite{Khalil2001}]
Let $(\tild A,B)$ be controllable and $(\tild A,E)$ be observable. Then, the $p \times p$ transfer function matrix $G(s)=E(sI-\tild A)^{-1}B$ is strictly positive real if and only if there exists a symmetric matrix $P>0$, matrices $L, W$, and a constant $\epsilon>0$ such that
\begin{equation}\label{LMI:KYP}
\begin{aligned}
P \tild A + \tild A^\top P &\leq - \epsilon I \\
PB &= E^\top 
\end{aligned}
\end{equation}

\end{Lemma}
 

Now, we assume system \eqref{sys:LTI} to be a structured system in the sense that it is composed out of $N$ subsystems in the form
\begin{equation}\label{sys:LTI_k}
\begin{aligned}
\dot x_k &= A_k x_k + \sum_{j \in \mathcal{N}_k} A_{kj} x_j + B_k u_k \\
y_k &=E_k x_k + \sum_{j \in \mathcal{N}_k} E_{kj} x_j ,
\end{aligned}
\end{equation}
where $x_k \in \mathbb{R}^{n_k}, \sum_{k=1}^N n_k =n$, and  $u_k, y_k \in \mathbb{R}^{q_k}$, $\sum_{k=1}^N q_k = q$. The interconnection topology is represented by the graph $\mathcal{G}=(\mathcal{V}, \mathcal{E})$. Let  $\widetilde{G}(s)$ be defined as the transfer matrix from $U(s)=[u_1(s)^\top, ..., u_N(s)^\top]^\top$ to 
$Y(s)=[y_1(s)^\top, ..., y_N(s)^\top]^\top$. 

\begin{Remark}
Later, in the design procedure for the distributed estimators, we will refer the estimators to this class of interconnected systems, and moreover, the design conditions given by the distributed version of the KYP-Lemma show the reason for the conservativeness of the intuitive approach presented above.
\end{Remark}

The following theorem delivers a sufficient condition for $\widetilde{G}(s)$ being strictly positive real. In particular, instead of solving \eqref{LMI:KYP} for the global system, the 
equations can be decomposed into local subproblems.

\begin{Theorem}[Distributed KYP-Lemma]\label{Th:DKYP}
The $p\times p$ transfer function matrix $\widetilde{G}(s)$ is strictly positive real if there exist symmetric $n_k \times n_k$ matrices $P_k>0, k=1,...,N$, $n_k \times n_j$ matrices $P_{kj}=P_{jk}^\top$ for $(k,j)\in\mathcal{E}$, and constants $\epsilon, \pi_1,...,\pi_N >0$ such that for all $k=1,...,N$, it holds that
\begin{equation*}
\begin{aligned}
&\underbrace{\left[\begin{array}{c;{2pt/2pt}cccc}
Q_k(k,k)  & Q_k(k,j_1)  & Q_k(k,j_2) & \hdots & Q_k(k,j_{p_k})   \\
\hdashline[2pt/2pt] \vspace{-0.2cm} \\
*    & Q_k(j_1, j_1) & Q_k(j_1, j_2) & \hdots &  Q_k(j_1, j_{p_k}) \\
* &  * & Q_k(j_2, j_2) &  \hdots & Q_k(j_1, j_{p_k}) \\
* &  * & * & \ddots & \vdots \\
* &  * & * & * & Q_k(j_{p_k},j_{p_k}) \\
\end{array}\right]}_{Q_k}
\end{aligned}
\end{equation*}
\begin{equation}\label{LMI:DKYP}
\begin{aligned}
&\!\!+\!\!\underbrace{\left[\begin{array}{c;{2pt/2pt}cccc}
p_k \pi_k P_k \!+\! \epsilon I \!+\! \overline W_k &   &  & 0  &    \\
\hdashline[2pt/2pt]
    & \!\!- \pi_{j_1} P_{j_1} &  & 0 &   \\
 &   & \!\!\!\!\!\!- \pi_{j_2} P_{j_2} &   &  \\
0 &  0 &  & \ddots &  \\
 &   &  &  & \!\!\!\!\!\!- \pi_{j_{p_k}} P_{j_{p_k}}\\
\end{array}\right]}_{S_k} \!\!\leq \! 0,
\end{aligned}
\end{equation}
with 
\begin{equation}\label{eq:DKYP_Q_elements}
\begin{aligned}
Q_k(k,k) &=P_k A_{k} + A_k^\top P_{k}  \\
Q_k(k,j) &=P_k A_{kj} + A_k^\top P_{kj}  \text{ for } j\in \mathcal{N}_k\\
Q_k(j_1,j_2) &=P_{kj_1}^\top A_{kj_2} + A_{kj_1}^\top P_{kj_2}  \text{ for } j_1, j_2\in \mathcal{N}_k
\end{aligned}
\end{equation}
and
\begin{align}
\label{eq:DKYP}
\left[\begin{array}{c}
P_k B_k \\
\hdashline[2pt/2pt] \vspace{-0.2cm} \\
P_{kj_1}^\top B_{k} \\
\vdots \\
P_{kj_{p_k}}^\top B_{k}
\end{array}\right] &=
\left[\begin{array}{c}
E_k^\top \\
\hdashline[2pt/2pt] \vspace{-0.2cm} \\
E_{kj_1}^\top\\
\vdots \\
E_{kj_{p_k}}^\top\\
\end{array}\right] \\
\sum_{j\in \mathcal{N}_k} \|P_k^{-1} P_{kj}\| &< 1. \label{eq:block-diag_dominance}
\end{align}
\end{Theorem}

$\overline W_k \geq 0$ in \eqref{LMI:DKYP} is a positive semi-definite matrix that can be used as a weighting matrix, e.g. to achieve performance guarantees. For the sake of proving Theorem \ref{Th:DKYP}, it can be assumed the $\overline W_k = 0$. Before the proof,
we first introduce following Lemma on block-diagonal dominant matrices.
\begin{Lemma}[\cite{Zhang2010}]\label{Lemma:block-diagonal-dominance}
Let the matrix $P$ be partitioned such that
\begin{equation}\label{eq:structured_P}
P = \begin{bmatrix}
P_1 & P_{12} & \hdots & P_{1N} \\
P_{21} & P_2 &  & \vdots \\
\vdots & & \ddots & \vdots \\
P_{N1} & \hdots & \hdots & P_{N}
\end{bmatrix},
\end{equation}
with $P_k \in \mathbb{R}^{n_k \times n_k}$, $P_k>0$ for all $k=1,...,N$, and $P_{kj}=0$ if $(v_k, v_j) \not\in \mathcal{E}$. If the reduced matrix $R=(r)_{ij}$ with the elements $r_{ij}=1$ for $i=j$ and $r_{ij}= -\|P_{ii}^{-1} P_{ij} \|$ for $i \neq j$ is strictly diagonal dominant, then for any eigenvalue $\lambda$ of $P$, it holds that $\lambda>0$.
\end{Lemma}

\begin{pf}[Theorem \ref{Th:DKYP}] 
Let there be matrices $P_k$ and $P_{kj}$ satisfying the design conditions of Theorem \ref{Th:DKYP}, which are \eqref{LMI:DKYP}, \eqref{eq:DKYP}, \eqref{eq:block-diag_dominance}.
Now, consider the matrix $P$ as defined in \eqref{eq:structured_P}, where $P_{kj}=0$ for $(k,j) \not\in \mathcal{E}$. With $P_k>0$ for $k=1,...,N$ and \eqref{eq:block-diag_dominance}, we have that the off-diagonal elements of the reduced matrix $R$ are all negative and it holds that $|\sum_{j \neq i} r_{ij}| <1$. With the diagonal elements of $R$ being $1$, this implies diagonal-dominance of $R$. Thus, we can apply Lemma \ref{Lemma:block-diagonal-dominance}, and obtain $P>0$. 
Now, we need to show that $P$ is a feasible solution to the centralized SPR-Lemma \eqref{LMI:KYP}. 
\begin{itemize}
\item As $B$ is a block-diagonal matrix with $B_1,...,B_N$ being the diagonal-blocks, we immediately have $PB=E^\top$ when applying \eqref{eq:DKYP}
\item Let $x=[x_1^\top,...,x_N^\top]^\top$ be any global state vector. Then we have
\vspace{-0.2cm}
\begin{align*}
x^\top P x =& \sum_{k=1}^N x_k^\top \left( P_k x_k + \sum_{j\in \mathcal{N}_k} P_{kj} x_j \right) \\
x^\top P\tild A x =& \sum_{k=1}^N x_k^\top P_k \left(  A_k x_k+ \sum_{j \in \mathcal{N}_k}A_{kj} x_j \right) \\
&+ \sum_{k=1}^N \sum_{j\in \mathcal{N}_k} x_k^\top P_{kj} \left(A_j x_j + \sum_{i=\mathcal{N}_j}A_{ji}x_i \right)
\end{align*}
\vspace{-0.6cm}
\end{itemize}
Now, from the fact that $\mathcal{G}$ is undirected and $P_{kj}=P_{jk}^\top$ we observe that for every $(v_k, v_j) \in \mathcal{E}$ we have both $x_k^\top P_{kj} \dot{x}_j$ and $x_j^\top P_{jk} \dot{x}_k$ as parts of $x^\top P\tild A x$. Therefore, by replacing $x_k^\top P_{kj} \dot{x}_j$ with $x_j^\top P_{jk} \dot{x}_k$, we obtain
\vspace{-0.2cm}
\begin{equation}\label{eq:changing_indices}
\begin{aligned}
x^\top P\tild A x & = \sum_{k=1}^N x_k^\top P_k  \left( A_k x_k+ \sum_{j \in \mathcal{N}_k}A_{kj} x_j \right) \\
&+ \sum_{k=1}^N \sum_{j\in \mathcal{N}_k} x_j^\top P_{kj}^\top \left(A_k x_k + \sum_{i=\mathcal{N}_k}A_{ki}x_i \right).
\end{aligned}
\end{equation}

\vspace{-0.3cm}
Adding the transposed part $x^\top \tild A^\top P x$ results in the complete equation
\vspace{-0.2cm}
\begin{align*}
x^\top (P\tild A+ &\tild A^\top P) x = \sum_{k=1}^N x_k^\top (P_k A_k + A_k^\top P_k) x_k \\
&+ \sum_{k=1}^N\sum_{j \in \mathcal{N}_k} \left( x_k^\top P_k A_{kj} x_j +x_j^\top A_{kj}^\top P_k x_k \right) \\
&+ \sum_{k=1}^N \sum_{j\in \mathcal{N}_k} \left( x_j^\top P_{kj}^\top A_k x_k +x_k^\top A_k^\top P_{kj}  x_j \right) \\
&+ \sum_{k=1}^N \sum_{j\in \mathcal{N}_k} \sum_{i=\mathcal{N}_k} x_j^\top P_{kj}^\top A_{ki}x_i+x_i^\top A_{ki}^\top P_{kj}x_j,
\end{align*}
\vspace{-0.7cm}

where the right hand side can be further transformed to
\vspace{-0.2cm}
\begin{align*}
&\sum_{k=1}^N x_k^\top (\underbrace{P_k A_k + A_k^\top P_k}_{Q_k(k,k)}) x_k  \\
&+ \sum_{k=1}^N\sum_{j \in \mathcal{N}_k} x_k^\top (\underbrace{P_k A_{kj}+A_k^\top P_{kj}}_{Q_k(k,j)}) x_j \\
& +\sum_{k=1}^N\sum_{j \in \mathcal{N}_k} x_j^\top (\underbrace{P_{kj}^\top A_k+A_{kj}^\top P_k}_{Q_k^\top(k,j)}) x_k \\
&+ \sum_{k=1}^N \sum_{i,j\in \mathcal{N}_k} x_j^\top (\underbrace{P_{kj}^\top A_{ki}+A_{kj}^\top P_{ki}}_{Q_k(j,i)})x_i.
\end{align*}
\vspace{-0.7cm}

With \eqref{LMI:DKYP}, we now have
\begin{equation}\label{eq:DKYP_Lie_derivative_end}
\begin{aligned}
&x^\top (P\tild A+ \tild A^\top P) x \leq \\
&\quad \sum_{k=1}^N \left( -x_k^\top (p_k \pi_k P_k + \epsilon I+ \overline W_k) x_k + \sum_{j \in \mathcal{N}_k} x_j^\top \pi_j P_j x_j \right) \\
&x^\top (P\tild A+ \tild A^\top P) x \leq -\epsilon x^\top x - \sum_{k=1}^N x_k^\top \overline{W}_k x_k  
\end{aligned}
\end{equation}
and therefore, $P$ satisfies \eqref{LMI:KYP}.


\end{pf}

The sufficient conditions derived in Theorem \ref{Th:DKYP} lead to a set of $N$ coupled LMIs and $N$ equality constraints. For instance, if $100$ subsystems \eqref{sys:LTI_k} with dimension $10$ are interconnected in a ring-type topology $(v_i, v_{i+1}) \in \mathcal{E}$,  
\eqref{LMI:DKYP} involves $100$ LMIs with dimension $30 \times 30$. In particular, those LMIs are amendable to parallel computing algorithms. Similar technique can be applied as presented in \cite{Wu2015b}.

\begin{Remark}
In the design conditions of Theorem \ref{Th:DKYP}, \eqref{eq:block-diag_dominance} represents a block-diagonal dominance condition, which is used to ensure positive definiteness of the Lyapunov-function. This inequality can also be replaced by the additional LMI
\begin{equation}\label{LMI:diagonal-dominance}
\begin{aligned}
&\left[\begin{array}{c;{2pt/2pt}ccc}
\frac{1}{1+p_k}P_k  & \frac{1}{2}P_{kj_1} & \hdots & \frac{1}{2}P_{k,j_{p_k}}   \\
\hdashline[2pt/2pt] \vspace{-0.2cm} \\
*    & \frac{1}{1+p_{j_1}}P_{j_1} &  &  0 \\
* &   & \ddots &  \\
* &  0 &  & \frac{1}{1+p_{j_{p_k}}}P_{j_{p_k}} \\
\end{array}\right]>0,
\end{aligned}
\end{equation}
which may be easier to implement numerically. In case there is no exact knowledge about the individual degrees of the neighbors, $p_{j_1} ... p_{j_{p_k}}$ in \eqref{LMI:diagonal-dominance}, it suffices to replace the degrees with upper bounds.
\end{Remark}

\begin{Remark}
The conditions of the Distributed KYP, Theorem \ref{Th:DKYP}, are sufficient conditions and thus there is a certain amount of conservativeness. However, conservativeness is expected to be small as it is only introduced by the coupling terms in the second line of \eqref{LMI:DKYP} and the assumption of diagonal dominance \eqref{LMI:diagonal-dominance} of $P$. A numerical example is shown later in the paper.
\end{Remark}

Concerning the interconnection topology $\mathcal{G}$, we can derive the following result for the case of identical $B_k$.

\begin{Corollary}
Suppose system \eqref{sys:LTI} is composed out of $N$ subsystems \eqref{sys:LTI_k} where the interconnection topology is represented by a $\mathcal{G}$. Let the $p \times p$ transfer function matrix $G(s)=E(sI-\tild A)^{-1}$ satisfy \eqref{LMI:KYP} and let $B_k = B_j \neq 0$ for two subsystems $k, j$. If $E_{kj} \neq 0$ and $E_{jk} = 0$, then it holds that $E_{kj}B_k=0$.
\vspace{-0.2cm}

\begin{pf}
Let $P$ be partitioned as shown in \eqref{eq:structured_P}. From symmetry of $P$, we have $P_{kj}=P_{jk}^\top$. Now, let $E_{jk} = 0$, then with \eqref{eq:DKYP} we have 
\vspace{-0.2cm}
\begin{equation*}
B_k^\top E^\top_{kj}=B_k^\top P_{kj}^\top B_k = B_j^\top P_{jk} B_k  = E_{jk} B_k = 0.
\end{equation*} 
\end{pf}
\end{Corollary}
\vspace{-0.4cm}


This corollary considers a special case of \eqref{sys:LTI_k}, which applies to the distributed estimator design presented in the next section. In fact, $E_{kj}$ is a design parameter for the distributed estimators, if $(v_j, v_k) \in \mathcal{E}$. This corollary shows that in the case of a directed graph, where $E_{kj}$ is a design parameter but $E_{jk}=0$, the choice of $E_{kj}$ is severely restrained. 

The LMIs \eqref{LMI:DKYP} give an analysis method for showing the SPR property for a network of interconnected systems by solving smaller feasibility problems. In particular, the individual feasibility problems only take local variables into account, which is essential for the distributed character of the problem. 
In the next section, distributed estimators will be designed, but since they are subject to disturbances, additional rows and columns will be added to \eqref{LMI:DKYP}.

\section{Distributed estimator design}\label{Section.Design}

%
%
%
%
%
%

\subsection{Estimator setup}

The estimator dynamics are proposed as
\begin{equation}\label{sys:nonlin_estimator}
\begin{aligned}
\dot{\hat{x}}_k =& A \hat{x}_k + B_\phi \hat{\phi}_k + B_\theta \theta(\tild H \hat x_k)+g(u) + L_k(y_k- C_k \hat{x}_k) \\
&+\sum_{j \in \mathcal{N}_k} K_{kj} (\hat{x}_j- \hat{x}_k) \\
\hat{\phi}_k =& \phi  \underbrace{\left( H \hat{x}_k + \tild L_k(y_k - C_k \hat{x}_k) +\sum_{j \in\mathcal{N}_k} \tild K_{kj} (\hat{x}_j- \hat{x}_k)\right)}_{v_k} 
\end{aligned}
\end{equation}

with initial condition $\hat{x}^{(k)}_0 $. The filter gains to be designed are $L_k, \tild L_k, K_{kj}$, and $\tild K_{kj}$, which are real matrices of suitable dimension. 

We can now particularize Problem 1 with respect to the proposed estimator dynamics.

\textbf{Problem 1':}
For all $k=1,...,N$ determine the estimator gains $L_k, \tild L_k, K_{kj}$, and $\tild K_{kj}$ in \eqref{sys:nonlin_estimator} such that the two properties of Problem 1 are satisfied simultaneously. 



\subsection{Filter gains design}

For the estimator error, we obtain with \eqref{sys:nonlinear} and \eqref{sys:nonlin_estimator} that
\begin{equation}\label{eq:e_k_dot}
\begin{aligned}
\dot{e}_k =& A e_k + B_\phi (\phi(Hx)-\hat{\phi}_k) +  B_\theta(\theta(\tild H x)-\theta(\tild H \hat{x}_k))\\
&- L_k C_k e_k + \sum_{j \in \mathcal{N}_k} K_{kj} (e_j- e_k) + B_w w\\
=& (A-L_k C_k- \sum_{j \in \mathcal{N}_k} K_{kj}) e_k  + \sum_{j \in \mathcal{N}_k} K_{kj} e_j + B_w w \\
&+ B_\phi (\phi(Hx)-\phi(v_k)) + B_\theta (\theta(\tild H x)-\theta(\tild H \hat{x}_k)).
\end{aligned}
\end{equation}
Following the argument from \cite{Arcak2001}, we replace the nonlinearities $\phi(Hx)-\phi(v_k)$ with the time-varying nonlinearities
\begin{equation}\label{eq:definition_psi}
\begin{aligned}
\psi_k(z_k, t) &= \phi(Hx)-\phi(v_k) \\
z_k &= H e_k - \tild L_k C_k e_k + \sum_{j \in\mathcal{N}_k} \tild K_{kj} (e_j- e_k) \\
z_k &= (H - \tild L_k C_k - \sum_{j \in\mathcal{N}_k} \tild K_{kj} )e_k + \sum_{j \in\mathcal{N}_k} \tild K_{kj}e_j
\end{aligned}
\end{equation}
Note that due to the monotonicity of $\phi(\cdot)$ \eqref{eq:monoton_nonlinearity}, $\psi_k(z_k, t)$ satisfies the sector property 
\begin{equation}\label{eq:sector_property}
z_k^\top \psi_k(z_k, t) \geq 0
\end{equation}
With this property, we are ready to present the main result, which delivers a design method for the distributed filter gains. 
\begin{Theorem}[$\mathcal{H}_\infty$-performance]\label{Th:nonlinear_performance}
Consider a nonlinear system \eqref{sys:nonlinear}. Define the following matrices
\begin{equation}\label{eq:redefining_matrices}
\begin{aligned}
A_k &= A-L_k C_k- \sum_{j \in \mathcal{N}_k} K_{kj}  \\ 
A_{kj} &= K_{kj} \\
E_k &= H - \tild L_k C_k - \sum_{j \in\mathcal{N}_k} \tild K_{kj} \\
E_{kj} &= \tild K_{kj} \\
B_k &= -B_\phi, \quad k=1,...,N
\end{aligned}
\end{equation}
and let the collection of matrices
$P_k, P_{kj}, L_k, \tild L_k, K_{kj}, \tild K_{kj}, k=1,...,N$, be a solution to the matrix inequalities

\vspace{-0.3cm}

\begin{equation}\label{BMI:circle_criterion_performance}
\left[\begin{array}{ccc;{2pt/2pt}cc}
&& &  P_k B_\theta & P_k B_w   \\
&Q_k + S_k & & P_{kj_1}^\top B_\theta & P_{kj_1}^\top B_w \\
&&&	 \vdots &  \vdots \\
&&&	 P_{kj_{p_k}}^\top B_\theta & P_{kj_{p_k}}^\top B_w \\
\hdashline[2pt/2pt]
*&*&* & -\tau^2 I & 0 \\
*&*&* & 0 &  -\gamma^2 I
\end{array}\right]<0,
\end{equation}
the distributed SPR-equations \eqref{eq:DKYP}, and the LMIs \eqref{LMI:diagonal-dominance} for $k=1,...,N$, where $Q_k, S_k$ are defined in \eqref{LMI:DKYP}, and $\overline W_k = W_k+ \tild H^\top \tild H$.

Then, the estimators \eqref{sys:nonlin_estimator} are a solution to Problem 1 in the sense of \eqref{eq:hinf-performance}, with performance parameter $\gamma$.

%

\begin{pf}

We use the Lyapunov function candidate
\begin{equation*}
\begin{aligned}
V(e)&= e^\top P e =\sum_{k=1}^N  \left( e_k^\top P_k e_k + \sum_{j\in \mathcal{N}_k} e_k^\top P_{kj} e_j \right).
\end{aligned}
\end{equation*}
With \eqref{eq:e_k_dot} and \eqref{eq:definition_psi}, the derivatives of $e_k$ can be reformulated to
\begin{equation}
\begin{aligned}
\dot e_k = & A_k e_k  + \sum_{j \in \mathcal{N}_k} A_{kj} e_j + B_w w \\
&- B_k \psi_k(z_k,t) + B_\theta (\underbrace{\theta(\tild H x)-\theta(\tild H \hat{x}_k)}_{\Delta_k}),
\end{aligned}
\end{equation}
and in addition, as \eqref{LMI:DKYP} is satisfied by \eqref{BMI:circle_criterion_performance}, we know that  \eqref{eq:DKYP_Lie_derivative_end} holds which ensures that
\vspace{-0.3cm}
\begin{equation}\label{eq:V_dot_stability}
\begin{aligned}
e^\top (P \tild A+ \tild A^\top P) e &= \sum_{k=1}^N e_k^\top Q_k  e_k < -\sum_{k=1}^N e_k^\top \overline W_k e_k.  
\end{aligned}
\end{equation}
For the Lie-derivative of $V(e)$, applying \eqref{eq:e_k_dot}, \eqref{eq:V_dot_stability}, and the same change of index as in \eqref{eq:changing_indices} leads to
\begin{equation*}
\begin{aligned}
\dot V(e)=&\sum_{k=1}^N \left( e_k^\top P_k \dot e_k + \dot e_k^\top P_k e_k \right) \\
& + \sum_{k=1}^N \sum_{j\in \mathcal{N}_k} \left( e_j^\top P_{kj}^\top \dot e_k + \dot e_k^\top P_{kj} e_j \right) 
\\
=&  \sum_{k=1}^N \left( e_k^\top P_k + \sum_{j \in \mathcal{N}_k} e_j^\top P_{kj}^\top \right) \dot e_k \\
&+ \sum_{k=1}^N \dot e_k^\top \left( P_k e_k + \sum_{j \in \mathcal{N}_k} P_{kj} e_j \right)\\
=&  \sum_{k=1}^N e_k^\top Q_k e_k \\
& - 2 \sum_{k=1}^N \left( e_k^\top P_k + \sum_{j \in \mathcal{N}_k} e_j^\top P_{kj}^\top \right) B_k \psi_k(z_k,t)\\
& +2 \sum_{k=1}^N \left( e_k^\top P_k + \sum_{j \in \mathcal{N}_k} e_j^\top P_{kj}^\top \right) (B_w w +B_\theta \Delta_k)
\end{aligned}
\end{equation*}
With \eqref{BMI:circle_criterion_performance} and the sector property \eqref{eq:sector_property} this further simplifies to
\vspace{-0.3cm}
\begin{equation}
\begin{aligned}
\dot{V}(e)<& \sum_{k=1}^N \left( -e_k^\top \overline W_k e_k + \tau^2 \Delta_k^\top \Delta_k + \gamma^2 w^\top w \right) \\
&- \sum_{k=1}^N \left( e_k^\top P_k B_k + \sum_{j \in \mathcal{N}_k} e_j^\top P_{kj}^\top B_k \right)  \psi_k(z_k,t) \\
= &\sum_{k=1}^N \left( -e_k^\top \overline W_k e_k + \tau^2 \Delta_k^\top \Delta_k + \gamma^2 w^\top w \right) \\
&- \sum_{k=1}^N \underbrace{ \left( e_k^\top E_k^\top + \sum_{j \in \mathcal{N}_k} e_j^\top E_{kj}^\top \right)}_{z_k^\top}   \psi_k(z_k,t)
\end{aligned}
\end{equation}

With the quadratic constraint \eqref{eq:qc-nonlinearity} on the nonlinearity $\theta$, we have $\tau^2 \Delta_k^\top \Delta_k< e_k^\top \tild H^\top \tild H e_k$. Then, due to the definition $\overline W_k = W_k+ \tild H^\top \tild H$ and the sector property of $\psi_k$ \eqref{eq:sector_property}, the Lie-derivative of $V(e)$ finally is
\begin{equation}
\begin{aligned}
\dot V(e) &< \sum_{k=1}^N \left( -e_k^\top W_k e_k + \gamma^2 w^\top w \right).
\end{aligned}
\end{equation}
Integrating over $(0,\infty)$ now yields the desired $\mathcal{H}_\infty$-performance
\begin{equation}
\begin{aligned}
\int_{0}^\infty \dot V(e) dt+ \sum_{k=1}^N \int_{0}^\infty e_k^\top W_k e_k dt &<  \int_{0}^\infty N \gamma^2 w^\top w dt \\
\sum_{k=1}^N \int_{0}^\infty e_k^\top W_k e_k dt &<  \gamma^2 \| w \|^2_{\mathcal{L}_2} + I_0,
\end{aligned}
\end{equation}
for $I_0 = V(e(0))$.

\end{pf}
\end{Theorem}


%
%

\vspace{-0.7cm}

Theorem \ref{Th:nonlinear_performance} gives us sufficient conditions for designing distributed estimators that satisfy the distributed KYP-Lemma from \ref{Th:DKYP} and moreover guarantees robust performance with respect to input disturbances. When substituting \eqref{eq:redefining_matrices}, however,one can easily see that the matrix inequalities \eqref{Th:DKYP} are not linear in the solution variables $L_k, K_{kj}, P_k, P_{kj}$ for $k=1,...,N$ and $j \in \mathcal{N}_k$. In the intuitive approach, when there are no off-diagonal blocks $P_{kj}$, a simple substitution $G_k = P_k L_k$ and $F_k = P_k K_k$
suffices to turn the matrix inequalities into LMIs. Since in the general case, off-diagonal blocks may be nonzero, we need to investigate in efficient solution strategies that can specifically find a suitable solution to the conditions of Theorem \ref{Th:DKYP}. 

\vspace{-0.3cm}

\subsection{Numerical calculation}

\vspace{-0.3cm}

Through the off-diagonal blocks, the problem becomes non-convex as discussed in many papers on decentralized control, e.g. \cite{Scherer2002}, \cite{Stankovic2007}, \cite{Swigart2010}. Due to this non-convexity, there is no general solution method available, but instead, alternative design methods are required. While Youla-Parametrization as in \cite{Scherer2002}, \cite{Swigart2010} is not suitable for the class of interconnection graphs under consideration, and a robustness argument as used in \cite{Stankovic2007} has proven as too conservative in the present context. In the following we will present a two-step solution strategy that has proven to be efficient of solving the matrix inequalities \eqref{BMI:circle_criterion_performance}.

\noindent\textbf{Step 1:}
Solve \vspace{-0.5cm}
\begin{equation}\label{optimization:Step1}
\begin{aligned}
& \min \sum_{k=1}^N \sum_{j \in \mathcal{N}_k} \| P_{k,j} \| \\
& \text{subject to } \eqref{eq:DKYP}, \eqref{LMI:diagonal-dominance},\eqref{BMI:circle_criterion_performance}, 
\end{aligned}
\end{equation}
where \eqref{BMI:circle_criterion_performance} is defined with
\begin{equation*}
\begin{aligned}
&Q_k=\left[\begin{array}{c;{2pt/2pt}cccc}
Q_k(k,k)  & Q_k(k,j_1)  & Q_k(k,j_2) & \hdots & Q_k(k,j_{p_k})   \\
\hdashline[2pt/2pt] 
*    & Q_k(j_1, j_1) & Q_k(j_1, j_2) & \hdots &  Q_k(j_1, j_{p_k}) \\
* &  * & Q_k(j_2, j_2) &  \hdots & Q_k(j_1, j_{p_k}) \\
* &  * & * & \ddots & \vdots \\
* &  * & * & * & Q_k(j_{p_k},j_{p_k}) \\
\end{array}\right]
\end{aligned}
\end{equation*}
with  \vspace{-0.3cm}
\begin{equation}\label{eq:DKYP_Q_elements2}
\begin{aligned}
Q_k(k,k) =& P_k A - G_k C_k - \sum_{l \in \mathcal{N}_k} F_{kl} \\
 &+ A^\top P_{k} - C_k^\top G_k^\top -  \sum_{l \in \mathcal{N}_k} F_{kl}^\top \\
Q_k(k,j) =& F_{kj} + A^\top P_{kj} - \lambda_k C_K^\top C_k P_{kj} - \lambda_k p_k P_{kj} \\
Q_k(j_1,j_2) =&\lambda_k P_{kj_1}^\top  + \lambda_k P_{kj_2}  \quad \text{ for } j, j_1, j_2\in \mathcal{N}_k
\end{aligned}
\end{equation}
and $S_k$ defined in \eqref{LMI:DKYP}. $G_k$ and $F_{kj}$ are matrix variables of suitable dimension and $\lambda_k, \pi_k, k=1,...,N$ are scalar parameters.

Through replacing the elements of $Q_k$ from \eqref{eq:DKYP_Q_elements} with \eqref{eq:DKYP_Q_elements2}, the matrix inequality \eqref{BMI:circle_criterion_performance} is turned into a LMI. Moreover, the minimization \eqref{optimization:Step1} can be executed in parallel fashion. The minimization of the off-diagonal blocks $P_{kj}$ ensures that they are only as large as needed for \eqref{eq:DKYP}. Then, the exact filter gains need to be calculated in the second step, where the feasibility is enhanced if the off-diagonal blocks are small.


\noindent\textbf{Step 2:}
Set $P_k, k=1,...,N$ and $P_{kj}, j\in \mathcal{N}_k$ as the results from Step 1 and solve \eqref{BMI:circle_criterion_performance} with the remaining variables. Optionally, $\gamma$ can be also defined as variable to be minimized.

An example where this 2 step approach is used will be given in the following. This two-step approach for computation has proven capable of solving numerous cases where the intuitive approach from Section 3 fails.

\section{Simulation example}
\begin{figure}[h!]
  \centering
 	\input{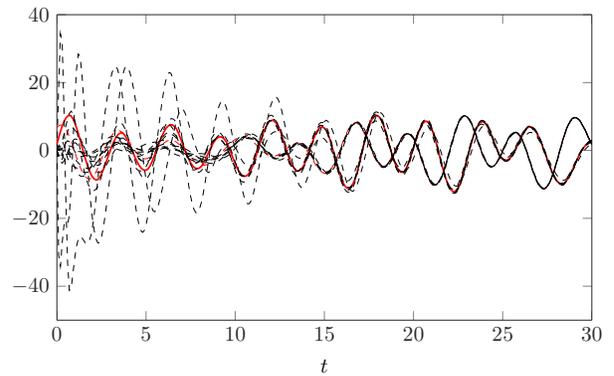}
 	\caption{Plots of $x_1$ and $x_2$. Red is the actual state, black are the estimates.}	
 	\label{fig:nonlinear_states}
 \end{figure}

\begin{figure}[h!]
  \centering
 	\input{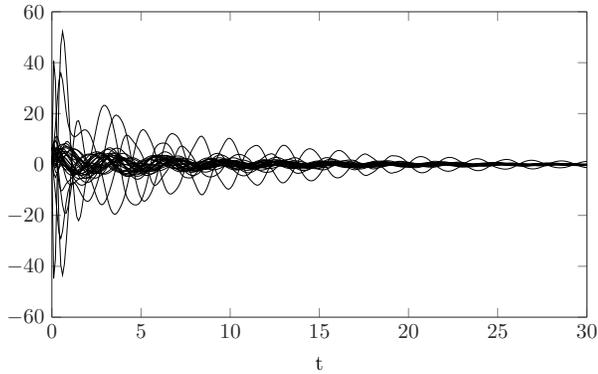}
 	\caption{Plots of the estimation error of all estimators.}	
 	\label{fig:nonlinear_error}
 \end{figure}

We consider our example from Section 3 \eqref{sys:example_6dim_oszi}, \eqref{sys:example_6dim_oszi_output}. Figure \ref{fig:nonlinear_states} and \ref{fig:nonlinear_error} show the simulation results after applying our 2-step approach with the parameters $\pi_k=0.1$ and $\lambda_k=1$ for all $k=1,...,N$, and the performance parameter $\gamma=4$. The nonlinearity is defined as $\phi(y)=\sqrt[3]{y}$.


\section{Conclusion}
In this paper, we discussed the extension of results from distributed estimation to nonlinear systems. While globally Lipschitz nonlinearities pose little problems, the Circle Criterion approach is far more challenging, requiring us to relax the usual assumption of a sum-of-squares Lyapunov-function. The new problem is non-convex, however, we presented an efficient solution algorithm which makes use of the exact structure of the problem, and is suitable for distributed calculation.

\section{Acknowledgment}

The authors would like to thank Prof. Hyungbo Shim, Prof. Valery Ugrinovskii, and Dr. Liron Allerhand for fruitful discussions.

\bibliographystyle{unsrt}
\bibliography{dkyp}

\end{document}